\documentclass[11pt, 
preprintnumbers,superscriptaddress,amsmath,amssymb,nofootinbib]{revtex4}
\usepackage[dvipdfmx]{graphicx}
\usepackage{dcolumn}
\usepackage{bm}
\usepackage{amssymb}
\usepackage{amsmath}
\usepackage{epsfig}    
\usepackage[dvipdfmx]{color}
\usepackage{slashed}
\usepackage{hhline}
\def\be{\begin{equation}}
\def\ee{\end{equation}}
\newcommand{\bea}{\begin{eqnarray}}
\newcommand{\eea}{\end{eqnarray}}
\newcommand{\nn}{\nonumber}

\numberwithin{equation}{section}

\begin{document}
    \rightline{KIAS-P19016, APCTP Pre2019-006}

\title{Muon anomalous magnetic moment, $Z$ boson decays, and collider physics in multi-charged particles}

\author{Takaaki Nomura}
\email{nomura@kias.re.kr}
\affiliation{School of Physics, KIAS, Seoul 02455, Republic of Korea}

\author{Hiroshi Okada}
\email{hiroshi.okada@apctp.org}
\affiliation{Asia Pacific Center for Theoretical Physics (APCTP) - Headquarters San 31, Hyoja-dong,
Nam-gu, Pohang 790-784, Korea}
\affiliation{Department of Physics, Pohang University of Science and Technology, Pohang 37673, Republic of Korea}

\date{\today}

\begin{abstract}
We analyze allowed region of muon anomalous magnetic moment (muon $g-2$), satisfying lepton flavor violations, $Z$ boson decays, and collider physics,
in a framework of multi-charged particles. Then we explore the typical size of the muon $g-2$, and discuss which mode dominantly affects muon $g-2$.
\end{abstract}
\maketitle

\newpage

\section{Introduction}
Muon anomalous magnetic moment (muon $g-2$) is one of the promising phenomenologies to confirm the new physics.
Therefore it still remains discrepancy between the standard model (SM) and new physics~\cite{Hagiwara:2011af};
\begin{align}
\Delta a_\mu=(26.1\pm8)\times10^{-10},\label{eq:damu}
\end{align}
where the $3.3\sigma$ deviation from the SM prediction with 
a positive value; recent theoretical analysis further indicates 3.7$\sigma$ deviation~\cite{Keshavarzi:2018mgv}.
Furthermore, several upcoming experiments such as Fermilab E989 \cite{e989} and J-PARC E34 \cite{jpark}
will provide the result with more precise manner. In theoretical point of view, several  mechanisms have been historically proposed through, e.g., gauge contributions~\cite{Altmannshofer:2014pba, Mohlabeng:2019vrz, Abdallah:2011ew}, Yukawa contributions at one-loop level~\cite{Lindner:2016bgg},  and Barr-Zee contributions~\cite{bz} at two-loop level. 
Especially, when one supposes the muon $g-2$ would be related to the other phenomenologies such as neutrino masses and dark matter candidate, Yukawa contributions at one-loop level would be likely to be promising candidates~\cite{Ma:2001mr, Okada:2013iba, Baek:2014awa, Okada:2014nsa, Okada:2014qsa, Okada:2015hia, Okada:2016rav, Nomura:2016rjf, Ko:2016sxg, Baek:2016kud, Nomura:2016ask, Lee:2017ekw, Chiang:2017tai, Das:2017ski, Nomura:2017ezy, Nomura:2017tzj, Cheung:2017kxb, Cheung:2018itc, Cai:2017jrq, Chakrabarty:2018qtt, CarcamoHernandez:2019xkb,Chen:2019nud,Nomura:2017ohi,Baumholzer:2018sfb}.  
In this case, one has to simultaneously satisfy several constraints of lepton flavor violations (LFVs) such as; $\ell_i\to \ell_j\gamma$, $\ell_i\to\ell_j\ell_k\bar\ell_\ell$($i,j,k,\ell=(e,\mu,\tau)$), and lepton flavor conserving(violating) $Z$ boson decays such as $Z\to\ell\bar\ell'$, $Z\to\nu\bar\nu'$~\cite{pdg}. 
Particularly, $\ell_\mu\to \ell_e\gamma$ gives the most stringent constraint, and the current branching ratio should be less than $4.2\times10^{-13}$~\cite{TheMEG:2016wtm}, and its future bound will reach at $6\times10^{-14}$~\cite{Renga:2018fpd}. Also $Z$ boson decays will be tested by a future experiment such as CEPC~\cite{cepc}.

In this paper, we introduce several multi-charged fields (bosons and fermions) with general $U(1)_Y$ hypercharges to get positive muon $g-2$, and we estimate the allowed region to satisfy all constraints of the muon $g-2$, LFVs, and $Z$ boson decays.
Also, we consider the constraint of collider physics, since multi-charged fields are severely restricted by the Large Hadron Collider (LHC).
We discuss the necessity of extra charged scalar in order to make exotic charged leptons decay into the SM particles and 
decay chains of exotic charged particles.
Then the signature of exotic charged particles are explored and we consider an allowed scenario accommodating muon $g-2$ and collider constraints.

This paper is organized as follows.
In Sec.~II, we review the model and formulate LFVs, muon $g-2$, $Z$ boson decays, and renormalization group for $g_Y$.
In Sec.~III, we estimate the allowed region for each $N$, comparing to collider physics.
We conclude in Sec.~IV.

\section{Model setup and Constraints with common part}

\begin{table}[t]
\begin{tabular}{|c|c|c|c||c|c|}
\hline\hline  
 &~$L_L$ ~&~$e_R$ ~&~$L'_{L/R}$ ~& ~$H$~ & ~$h^{+n}$ \\\hline 
$SU(2)_L$ & $\bm{2}$& $\bm{1}$& $\bm{2}$   & $\bm{2}$ & $\bm{1}$   \\\hline 
$U(1)_Y$   & $-\frac12$ & $-1$ & $-\frac{N}2$ & $\frac12$ & $\frac{N-1}2$     \\\hline
\end{tabular}
\caption{Charge assignments of fields under $SU(2)_L\times U(1)_Y$, where $n\equiv\frac{N-1}{2}$ with $N=3,5,\cdots$, and all the new fields are color singlet.}
\label{tab:1}
\end{table}

In our set up of the model, we introduce an isospin doublet fermion $L'_a\equiv[\psi^{-n}_a,\psi^{-n-1}_a]^T\ (a=1)$ for simplicity~\footnote{When $L'$ provides a flavor structure of neutrino mass matrix, we minimally need two families to satisfy the neutrino oscillation data. }, and a new
boson $h^{+n}$ with $n\equiv\frac{N-1}2$ ($N=3,5,\cdots$), as shown
in Table~\ref{tab:1}. Notice here that $N$ is defined by odd number, where $N=1$ is not considered because $L'_L$ cannot be discriminated from $L_L$.
The valid Lagrangian is given by
\begin{align}
-\mathcal{L}^n_{Y}  &= f_{ia} \bar L_{L_i} L'_{R_a} h^{n} + {\rm h.c.}\nn\\
&=f_{ia}[\bar \nu_{L_i} \psi^{-n}_a h^{n}+\bar\ell_i\psi^{-n-1}_a h^n]+ {\rm h.c.},
\label{Eq:lag-yukawa} 
\end{align}
where $i=1-3,\ a=1$ are generation indices.
The Yukawa Lagrangian $y_{\ell_{ii}}\bar L_{L_i}e_{R_i}H$
provides masses for the charged leptons {$(m_{\ell_i}\equiv y_{\ell_{ii}}v/\sqrt2$)} by developing a nonzero 
vacuum expectation value (VEV) of $H$, which is denoted by $\langle H\rangle\equiv v/\sqrt2$.
The exotic lepton $L'$ has vector-like mass and new scalar field $h^{\pm n}$ does not develop a VEV.
We denote mass of $L'$ and $h^{\pm n}$ by $m_\psi$ and $m_h$ respectively.

\subsection{Lepton flavor violations and muon anomalous magnetic moment}
\label{lfv-lu}
The Yukawa terms of ($f,g$) give rise to $\ell_i\to\ell_j\gamma$ processes 
at one-loop level.
 %
The branching ratio is given by
\begin{align}
B(\ell_i\to\ell_j \gamma)
\approx 
\frac{48\pi^3 \alpha_{\rm em}}{{G_{\rm F}^2 m_{\ell_i}^2} } C_{ij}\left( |a_{L_{ij}}|^2 + |a_{R_{ij}}|^2\right),
\end{align}
where $G_{\rm F}\approx1.166\times 10^{-5}$ GeV$^{-2}$ is the Fermi constant, 
$\alpha_{\rm em}(m_Z)\approx {1/128.9}$ is the 
fine-structure constant~\cite{pdg}, 
$C_{21}\approx1$, $C_{31}\approx0.1784$, and $C_{32}\approx0.1736$.
$a_{L/R}$ is formulated as 
\begin{align}
& a_{L_{ij}}
\approx
- m_{\ell_i} \sum_{a=1-3} \frac{f_{ja} f^\dag_{ai} }{(4\pi)^2} 
\left[ n F(\psi^{-n-1}_a , h^n) + (n+1) F(h^n,\psi^{-n-1}_a)\right], \label{eq:lfv-L}\\
&a_{R_{ij}}
\approx
- m_{\ell_j} \sum_{a=1-3} \frac{f_{ja} f^\dag_{ai} }{(4\pi)^2} 
\left[ n F(\psi^{-n-1}_a , h^n) + (n+1) F(h^n,\psi^{-n-1}_a)\right], \label{eq:lfv-R}\\
&F_{}(1,2)\approx \frac{(m_1^2-m_2^2)\{5 m_1^2m_2^2-m_2^4(1+3n)+m_1^4(2+3n)\}- 12m_1^2 m_2^2
\{-nm_2^2+(1+n)m_1^2\}
\ln\left[\frac{m_1}{m_2}\right]}
{12(m_1^2-m_2^2)^4},
\label{eq:lfv-lp}
\end{align} 
where $m_{\psi^{-n-1}}\equiv m_\psi$, and $m_{h^n}\equiv m_h$.
The current experimental upper bounds are given 
by~\cite{TheMEG:2016wtm, Aubert:2009ag}
  \begin{align}
  B(\mu\rightarrow e\gamma) &\leq4.2\times10^{-13}(6\times10^{-14}),\quad 
  B(\tau\rightarrow \mu\gamma)\leq4.4\times10^{-8}, \quad  
  B(\tau\rightarrow e\gamma) \leq3.3\times10^{-8}~,
 \label{expLFV}
 \end{align}
where parentheses of $\mu\to e\gamma$ is a future reach of MEG experiment~\cite{Renga:2018fpd}.

{\it The muon anomalous magnetic moment} ($\Delta a_\mu$): 
We can also estimate the muon anomalous magnetic moment through ${a_{L,R}}$, which is given by 
\begin{align}
\Delta a_\mu\approx -m_\mu (a_L+a_R)_{22}.\label{eq:damu}
\end{align}
The $3.3\sigma$ deviation from the SM prediction is  
$\Delta a_\mu=(26.1\pm8)\times10^{-10}$~\cite{Hagiwara:2011af} with 
a positive value.

\begin{figure}[tb]
\begin{center}
\includegraphics[width=100mm]{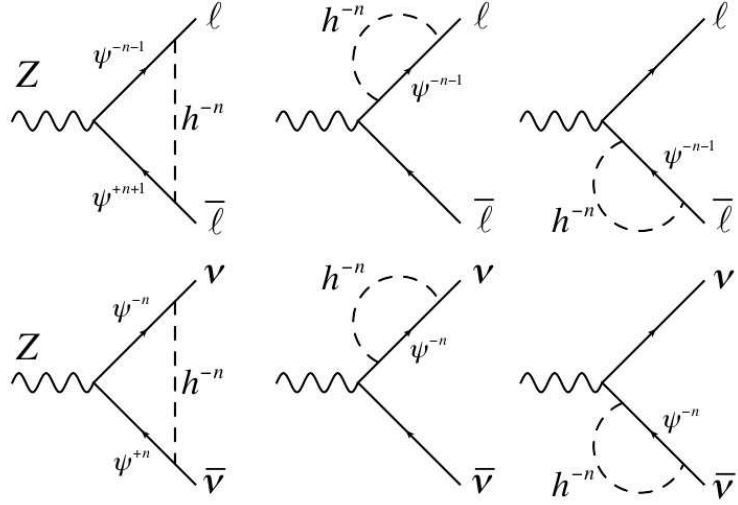}
\caption{Feynman diagrams for $Z\to \ell_i\bar\ell_j$ and $Z \to \nu_i \bar \nu_j$, where upper diagrams represent contribution to $Z \bar \ell  \ell$ while the down ones are for $Z \bar \nu \nu$.}
\label{fig:zto2ell}
\end{center}
\end{figure}

\subsection{Flavor-Conserving(Changing) Leptonic $Z$ Boson Decays}\label{subsec:Zll}
Here, we consider the $Z$ boson decay into two leptons through the Yukawa terms $f$ at one-loop level~\cite{Chiang:2017tai}.
Since some components of $f$ are expected to be large so as to obtain 
a sizable $\Delta a_\mu$, the experimental bounds on
$Z$ boson decays could be of concern at one loop level. 
First of all, the relevant Lagrangian is given by~\footnote{We neglect one-loop contributions in the SM.}
\begin{align}
{\cal L}&\sim
\frac{g_2}{c_w} \left[\bar\ell\gamma^\mu \left(-\frac12 P_L+s_W^2\right)\ell
+\frac12\bar\nu\gamma^\mu P_L\nu
 \right] Z_\mu\nn\\
&+ \frac{g_2}{c_w} \left[
 \left(-\frac12 P_L+n s_W^2\right)\bar\psi^{n}\gamma^\mu \psi^{-n}
+ \left(-\frac12 P_L+(n+1) s_W^2\right)\bar\psi^{n+1}\gamma^\mu \psi^{-n-1}
 \right] Z_\mu\nn\\
 & +in\frac{g_2  s_W^2}{c_W}(h^n\partial^\mu h^{-n} - h^{-n}\partial^\mu h^{n})Z_\mu,
\end{align}
where $s(c)_W\equiv\sin(\cos)\theta_W\sim0.23$ 
stands for the sine (cosine) of the Weinberg angle. 
The decay rate of the SM at tree level is then given by
\begin{align}
&{\rm \Gamma}(Z\to\ell^-_i\ell^+_j)_{SM}
\approx
\frac{m_Z}{12\pi} \frac{g_2^2}{c_W^2}  \left(s_W^4-\frac{s_W^2}2 + \frac18 \right)\delta_{ij},\\
&{\rm \Gamma}(Z\to\nu_i\bar\nu_j)_{SM}
\approx
\frac{m_Z}{96\pi} \frac{g_2^2}{c_W^2} \delta_{ij}.
\end{align}
Combining all the diagrams in Fig.~\ref{fig:zto2ell},
the ultraviolet divergence cancels out and only the finite part 
remains~\cite{Chiang:2017tai}.
The resulting form is given by
\begin{align}
&\Delta{\rm \Gamma}(Z\to\ell^-_i\ell^+_j)
\approx
\frac{m_Z}{12\pi} \frac{g_2^2}{c_W^2}
\left[ 
\frac{|B_{ij}^{\ell}|^2}{2} - {\rm Re}[A_{ij} (B^{\ell})^*_{ij}]
-\left(
-\frac{s_W^2}2 + \frac18\right)\delta_{ij}\right] \label{eq:Zll},\\
&\Delta{\rm \Gamma}(Z\to\nu_i\bar\nu_j)
\approx
\frac{m_Z}{24\pi} \frac{g_2^2}{c_W^2}
\left[ 
{|B_{ij}^{\nu}|^2} 
-\frac{\delta_{ij}}{4}\right] \label{eq:Znunu},
\end{align}
where 
\begin{align}
&A_{ij}\approx s^2_W\delta_{ij},
\quad B^{\ell}_{ij}\approx \frac{\delta_{ij}}{2} - \frac{f_{ia} f^\dag_{aj}}{(4\pi)^2} G^{\ell}(\psi,h),
\quad B^{\nu}_{ij}\approx \frac{\delta_{ij}}{2} + \frac{f_{ia} f^\dag_{aj}}{(4\pi)^2} G^{\nu}(\psi,h),
\\
&G^{\ell}(\psi,h)\approx-ns^2_W\left(-\frac12 +s_w^2\right) H_1(\psi,h)
-\left(-\frac12 +s_w^2\right)^2 H_2(\psi,h)
+\left(-\frac12 +(n+1) s_w^2\right) H_3(\psi,h),\\
&G^{\nu}(\psi,h)\approx-ns^2_W\left(-\frac12 +s_w^2\right) H_1(\psi,h)
- \frac12 H_2(\psi,h)
+\left(-\frac12 +n s_w^2\right) H_3(\psi,h),
\\
&H_1(1,2)= \frac{m_1^4-m_2^4+4 m_1^2 m_2^2\ln\left[\frac{m_2}{m_1}\right]}{2(m_1^2-m_2^2)^2},\\
&H_2(1,2)= \frac{m_2^4 - 4m_1^2 m_2^2 +3m_1^4 - 4 m_2^2(m_2^2-2m_1^2)\ln[m_2]-4m_1^4\ln[m_1]}{4(m_1^2-m_2^2)^2},\\
&H_3(1,2)=m_1^2\left( \frac{m_1^2-m_2^2 + 2 m_2^2\ln\left[\frac{m_2}{m_1}\right]}{(m_1^2-m_2^2)^2}\right).
\end{align}
Notice here that the upper index of $B$ represents $\psi\equiv \psi^{-n-1}$ for cahrged-lepton final state, while  $\psi\equiv \psi^{-n}$ for the neutrino final state.
One finds the branching ratio by dividing the total $Z$ decay width $\Gamma_{Z}^{\rm tot} = 2.4952 \pm 0.0023$~GeV~\cite{pdg}.
The current bounds on the lepton-flavor-(conserving)changing $Z$ boson decay 
branching ratios at 95 \% CL are given by \cite{pdg}:
\begin{align}
& \Delta {\rm BR}(Z\to {\rm Invisible})\approx  \sum_{i,j=1-3}\Delta {\rm BR}(Z\to\nu_i\bar\nu_j)< \pm5.5\times10^{-4} ,
\label{eq:zmt-con}\\
 & \Delta {\rm BR}(Z\to e^\pm e^\mp) < \pm4.2\times10^{-5} ~,\
 \Delta   {\rm BR}(Z\to \mu^\pm\mu^\mp) <  \pm6.6\times10^{-5} ~,\
 \Delta   {\rm BR}(Z\to \tau^\pm\tau^\mp) <  \pm8.3\times10^{-5} ~,\label{eq:zmt-con}\\
&    {\rm BR}(Z\to e^\pm\mu^\mp) < 7.5\times10^{-7} ~,\
  {\rm BR}(Z\to e^\pm\tau^\mp) < 9.8\times10^{-6} ~,\
  {\rm BR}(Z\to \mu^\pm\tau^\mp) < 1.2\times10^{-5} ~,\label{eq:zmt-cha}
\end{align}
where $\Delta {\rm BR}(Z\to f_i\bar f_j)$ ($i= j$) is defined by
\begin{align}
\Delta {\rm BR}(Z\to f_i \bar f_j)\approx 
\frac{{\rm \Gamma}(Z\to f_i \bar f_j)- {\rm \Gamma}(Z\to f_i \bar f_j)_{SM}}
{\Gamma_{Z}^{\rm tot}}.
\end{align}
We consider these constraints in our global analyses below.

\subsection{Beta function of $g_Y$}
\label{beta-func}
\begin{figure}[tb]
\begin{center}
\includegraphics[width=13cm]{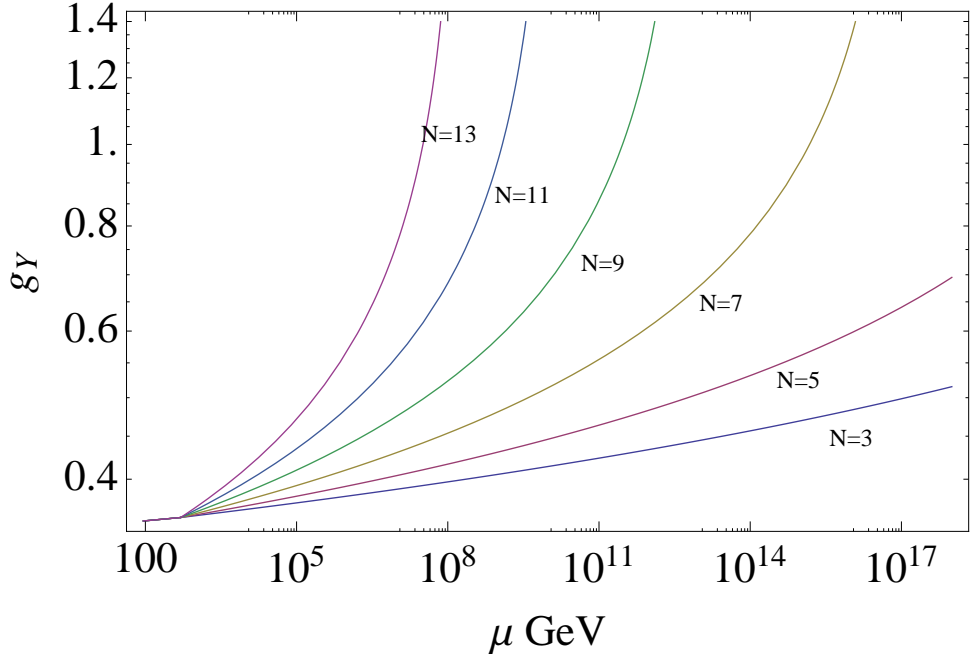}
\caption{The running of $g_Y$ in terms of a reference energy of $\mu$, depending on each of $N=3,5,7,9,11,13$.}
\label{fig:rge}
\end{center}\end{figure}

Here we estimate the effective energy scale by evaluating the Landau
pole for $g_Y$ in the presence of new exotic fields with nonzero
multiple hypercharges.  Each contribution of the new beta function 
of $g_Y$ from one 
$SU(2)_L$ doublet fermion with $-N/2$ hypercharge is given
by~\cite{Ko:2016sxg}
\begin{align}
\Delta b^f_Y={\frac{3}{5}\times}\frac{4}{3}\times\left(\frac{N}2\right)^2 \ .
\end{align}
Similarly, the contribution to the beta function from 
one $SU(2)_L$ singlet boson with $(N-1)/2$ hypercharge
is given by 
\begin{align}
\Delta b^b_Y={\frac{3}{5}\times}\frac{1}{3}\times\left(\frac{N-1}2\right)^2  ,
\end{align}
{where $3/5$ is the rescaled coefficient.}
Then one finds the energy evolution of the gauge coupling 
$g_Y$ as~\cite{Kanemura:2015bli}
\begin{align}
\frac{1}{g^2_Y(\mu)}&=\frac1{g_Y^2(m_{in.})}-\frac{b^{SM}_Y}{(4\pi)^2}\ln\left[\frac{\mu^2}{m_{in.}^2}\right]
-\theta(\mu-m_{thres.}) \frac{(\Delta b^f_Y+\Delta b^b_Y)}{(4\pi)^2}\ln\left[\frac{\mu^2}{m_{thres.}^2}\right]
,\label{eq:rge_gy}
\end{align}
where $\mu$ is a reference energy scale, and we assume that
$m_{in.}(=m_Z)<m_{thres.}=$500 GeV, where $m_{in.}$ $m_{thres.}$ are initial and threshold mass, respectively.
The resulting running of $g_Y(\mu)$ versus the scale $\mu$ 
is shown in Fig.~\ref{fig:rge} for 
each of $N=3,5,7,9,11,13$.

\section{Muon $g-2$ and physics of Each $N$}

In this section we estimate muon $g-2$ taking into account constraints from LFVs and Z decays 
and discuss constraint and prospect for collider physics in some number of $N$. 

In addition to the Yukawa interaction explaining muon $g-2$, 
we need extra particles and/or interactions to make exotic particles decay into SM ones.
Here we summarize extensions for the cases of $N=3$, $N=5$ and $N=7$ as follows.
\footnote{For $N$ is more than 7, model is rather complicated because some fields have to be introduced in order to make new fields into the SM.
Thus, we consider $N=3,5,7$.} \\
{\bf (1) $N=3$}: In this case, we have interaction term 
\begin{equation}
\mathcal{L}_{ex 1} = g_{ij} \bar L^c_{L_i} L_{L_j}h^+ + h.c. \ ,
\label{EQ:extra_int_1}
\end{equation}
without introducing extra particle.
Then all exotic particles can eventually decay into the SM particles. \\
{\bf (2) $N=5$}: in this case, we have interaction term 
\begin{equation}
\mathcal{L}_{ex 2} = g'_{ij} \bar e^c_{R_i} e_{R_j}h^{++} + h.c. \ ,
\label{EQ:extra_int_2}
\end{equation}
without introducing extra particle, 
and exotic particles can decay into the SM particles as in the $N=3$ case. 
Here it is also worthwhile mentioning the we can explain the active neutrino sector at two-loop level,
if both extra terms are introduced with extra doubly(singly) charged particle for $N=3(5)$ cases. This is called  Zee-Babu model~\cite{zee, babu}. \\
{\bf (3) $N=7$}: In this case, we need to introduce $h^{\pm}$ and $h^{\pm \pm}$ in addition to $h^{\pm \pm \pm}$ in order to make it decay into the SM particles.
We then have interactions $\mathcal{L}_{ex 1(2)}$ and new interaction in scalar potential:
\begin{equation}
V_{ex} = \mu_X h^{+++} h^{--} h^{-} + c.c. \ , 
\label{EQ:extra_int_3}
\end{equation}
with which triply charged scalar can decay into the SM particles through doubly and singly charged scalar decay by $\mathcal{L}_{ex 1(2)}$ interaction.
Note that new Yukawa interactions affect LFVs, muon $g-2$, and Z decays.
Especially, these terms contribute to the muon $g-2$ negatively.
Therefore, we require these terms are enough small to satisfy the sizable muon $g-2$. 

\begin{figure}[tb]
\begin{center}
\includegraphics[width=7cm]{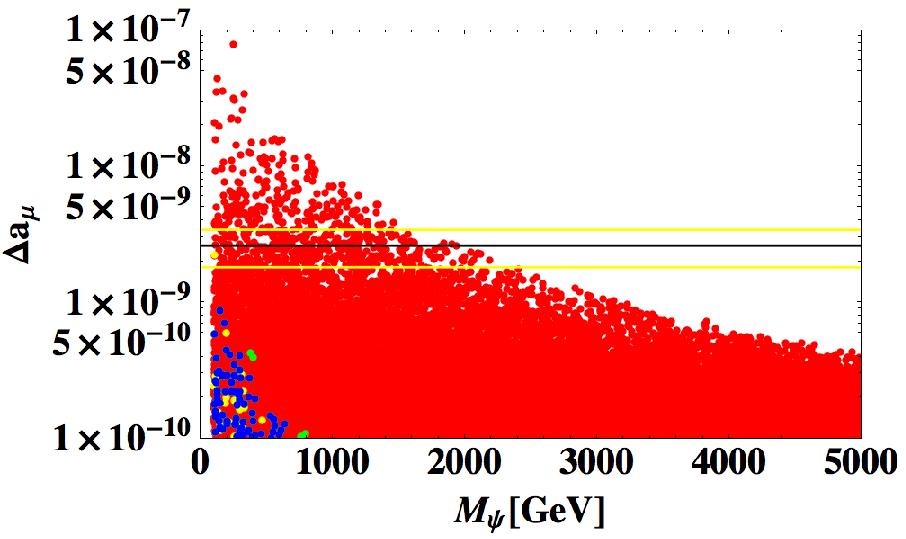}
 \includegraphics[width=7cm]{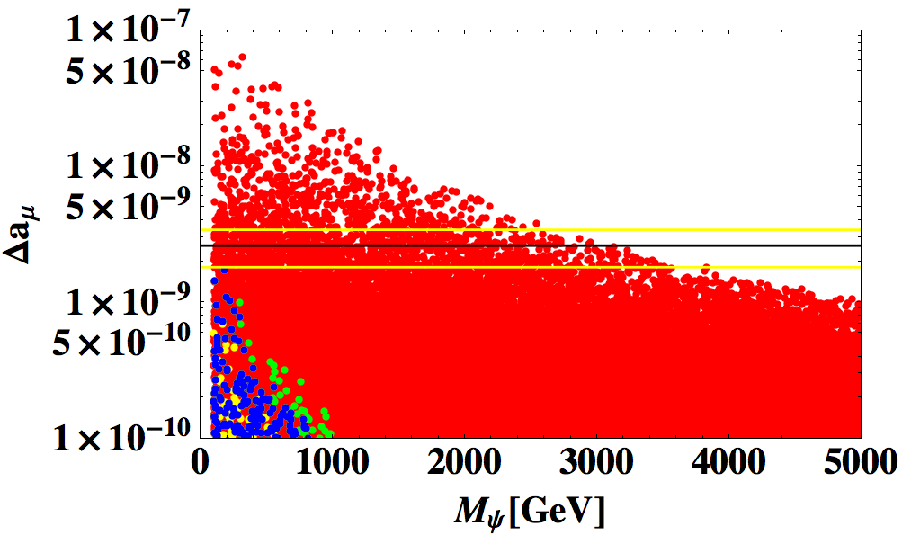}
 \includegraphics[width=7cm]{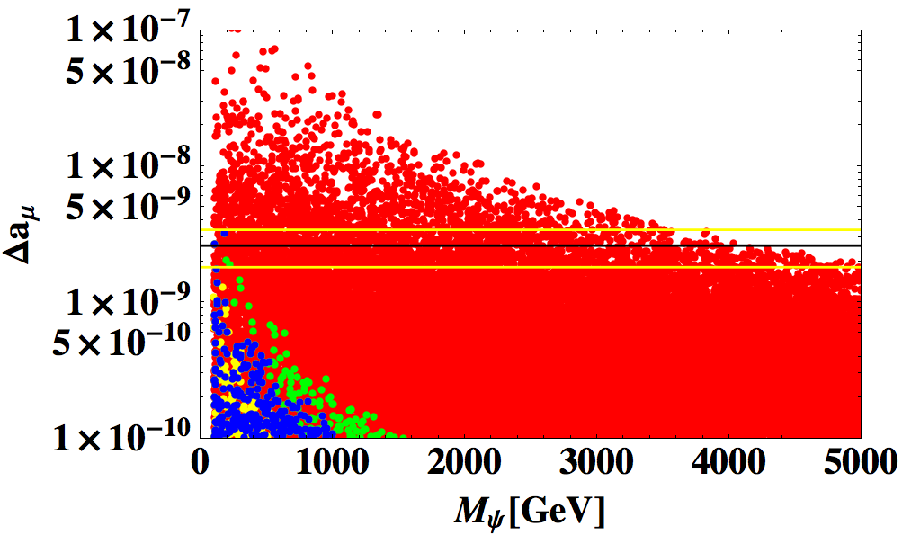}
\caption{Muon $g-2$ as a function of $L'$ mass obtained from parameter scan for $N=3$, $N=5$ and $N=7$  where red, green, yellow, and blue color points respectively correspond to those with $\ell_i \to \ell_j \gamma$ constraints, $\ell_i \to \ell_j \gamma$ plus $Z\to\nu_i\bar\nu_j$, $\ell_i \to \ell_j \gamma$ plus $Z\to\mu\bar\mu$, and  $\ell_i \to \ell_j \gamma$ plus all of the $Z\to f_i\bar f_j$.}
\label{fig:mg2-I}
\end{center}\end{figure}

\begin{figure}[tb]
\begin{center}
\includegraphics[width=7cm]{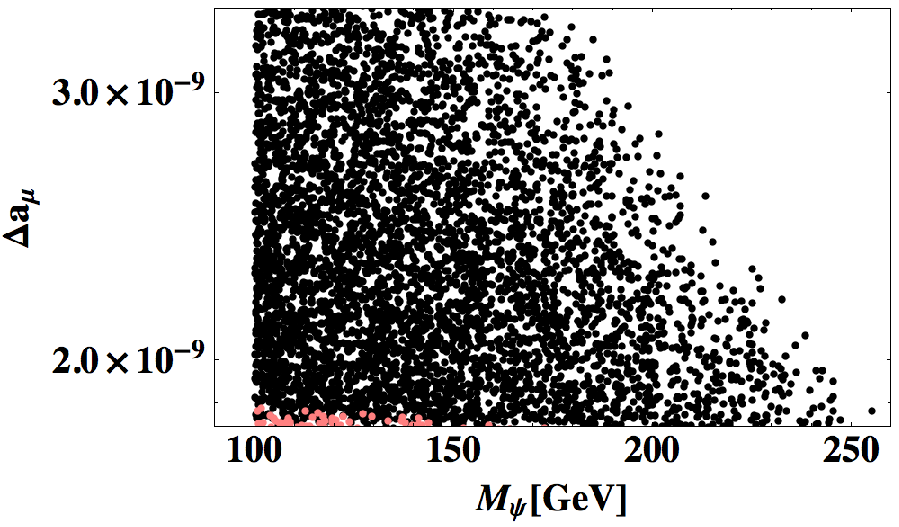}
 \includegraphics[width=7cm]{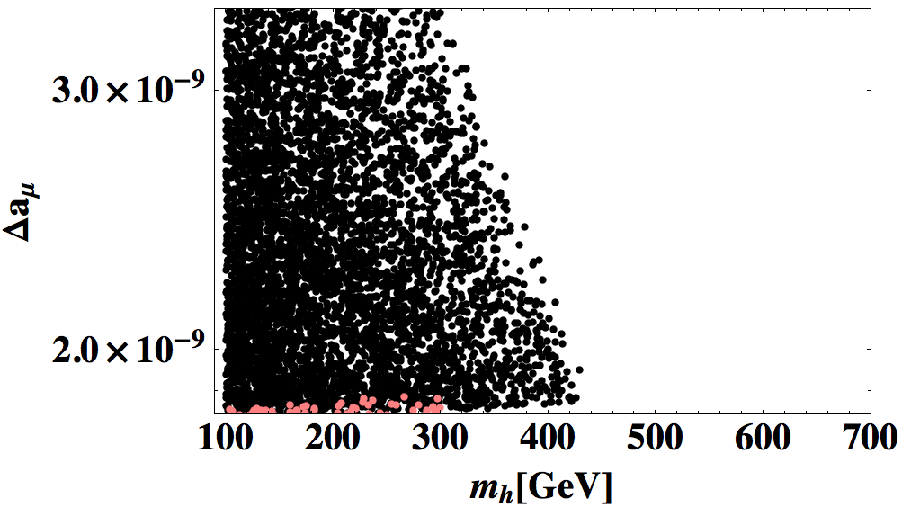}
\caption{Muon $g-2$ as a function of $L'$ and $h^{n}$ masses (left and right plots) obtained from parameter scan imposing all the constraints as discussed in Fig.~\ref{fig:mg2-I}, where black and pink points respectively correspond to cases of $N=5$ and $N=7$.
Note here that there are not any allowed points for $N=3$.}
\label{fig:mg2-II}
\end{center}\end{figure}

\subsection{Muon $g-2$ and flavor constraints for each case}

In this subsection, we scan Yukawa coupling in Eq.~(\ref{Eq:lag-yukawa}) and estimate muon $g-2$ taking into account constraints from LFV charged lepton decay as well as 
$Z \to \ell_i^+ \ell^-_j$ processes discussed in previous section.
Here we universally scan $f_{i1}$ in the range of 
\begin{equation}
f_{i1} \in [10^{-6}, \sqrt{4 \pi}],
\end{equation}
where the upper bound is requirement from perturbativity.
Firstly we take wide mass range of $\{m_\psi, m_h \} \in [100, 5000]$ GeV in our parameter scan where $m_\psi$ and $m_h$ are respectively mass of $L'$ and $h^n$.
In Fig.~\ref{fig:mg2-I}, we show the value of muon $g-2$ as a function of exotic lepton mass for $N=3$, $N=5$ and $N=7$ where red, green, yellow, and blue color points respectively correspond to those with $\ell_i \to \ell_j \gamma$ constraints, $\ell_i \to \ell_j \gamma$ plus $Z\to\nu_i\bar\nu_j$, $\ell_i \to \ell_j \gamma$ plus $Z\to\mu\bar\mu$, and  $\ell_i \to \ell_j \gamma$ plus all of the $Z\to f_i\bar f_j$.
We see that $Z\to\mu\bar\mu$ and $Z\to\nu_i\bar\nu_j$ constraints severely exclude the parameter region, and exotic particle masses are preferred to be relatively light as $m_{\psi, h} \lesssim 500$ GeV.
Then we focus on light mass region which can accommodate with muon $g-2$.
The left and right plots in Fig.~\ref{fig:mg2-II} show the value of muon $g-2$ as a function of $L'$ and $h^n$ masses respectively imposing all the constraints as discussed in Fig.~\ref{fig:mg2-I}, where black and pink points respectively correspond to cases of $N=5$ and $N=7$.
Furthermore we show contour plot for $\Delta a_\mu$ and $\Delta BR_{\mu \mu} \equiv \Delta BR (Z \to \mu^+ \mu^-)$ on $\{M(=m_h = m_\psi), f_{21} \}$ plane where we take only $f_{21}$ to be non-zero and other $f_{ij}$ to be zero. In the plots, the (light-)yellow region is $(2 \sigma) 1 \sigma$ region for muon $g-2$ and shaded region is excluded by $\Delta BR (Z \to \mu^+ \mu^-)$.
Thus one find that the mass scale is constrained by $\Delta BR_{\mu \mu}$ even if only $f_{21}$ is non-zero.
We thus find that $L'$ mass should be relatively light as $m_\psi \sim (150, 200, 250 )$ GeV for $N=(3, 5,7)$ to explain muon $g-2$ within $1 \sigma$ while charged scalar mass $m_h$ can be heavier than $m_\psi$.

  \begin{figure}[tb]
\begin{center}
\includegraphics[width=7cm]{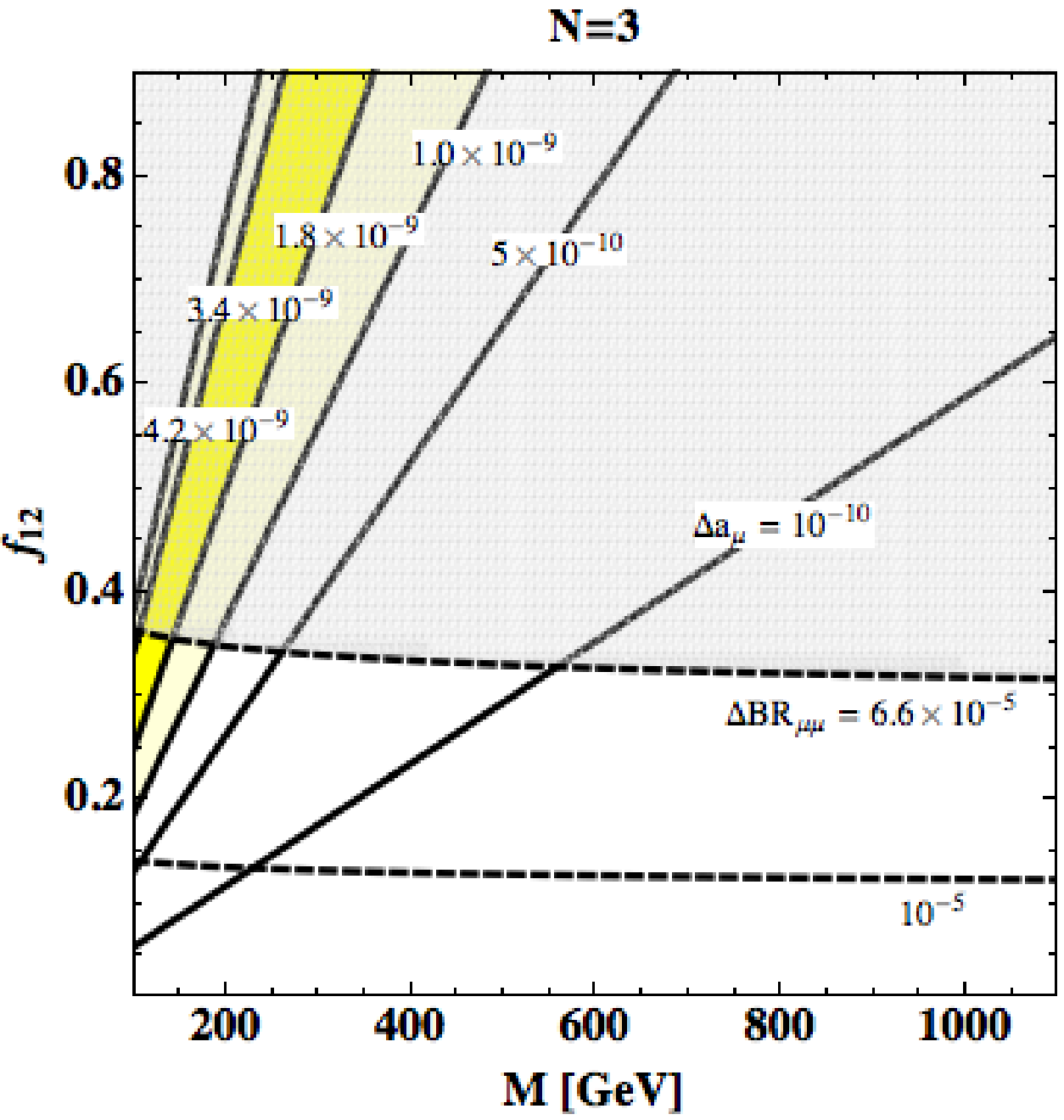}
 \includegraphics[width=7cm]{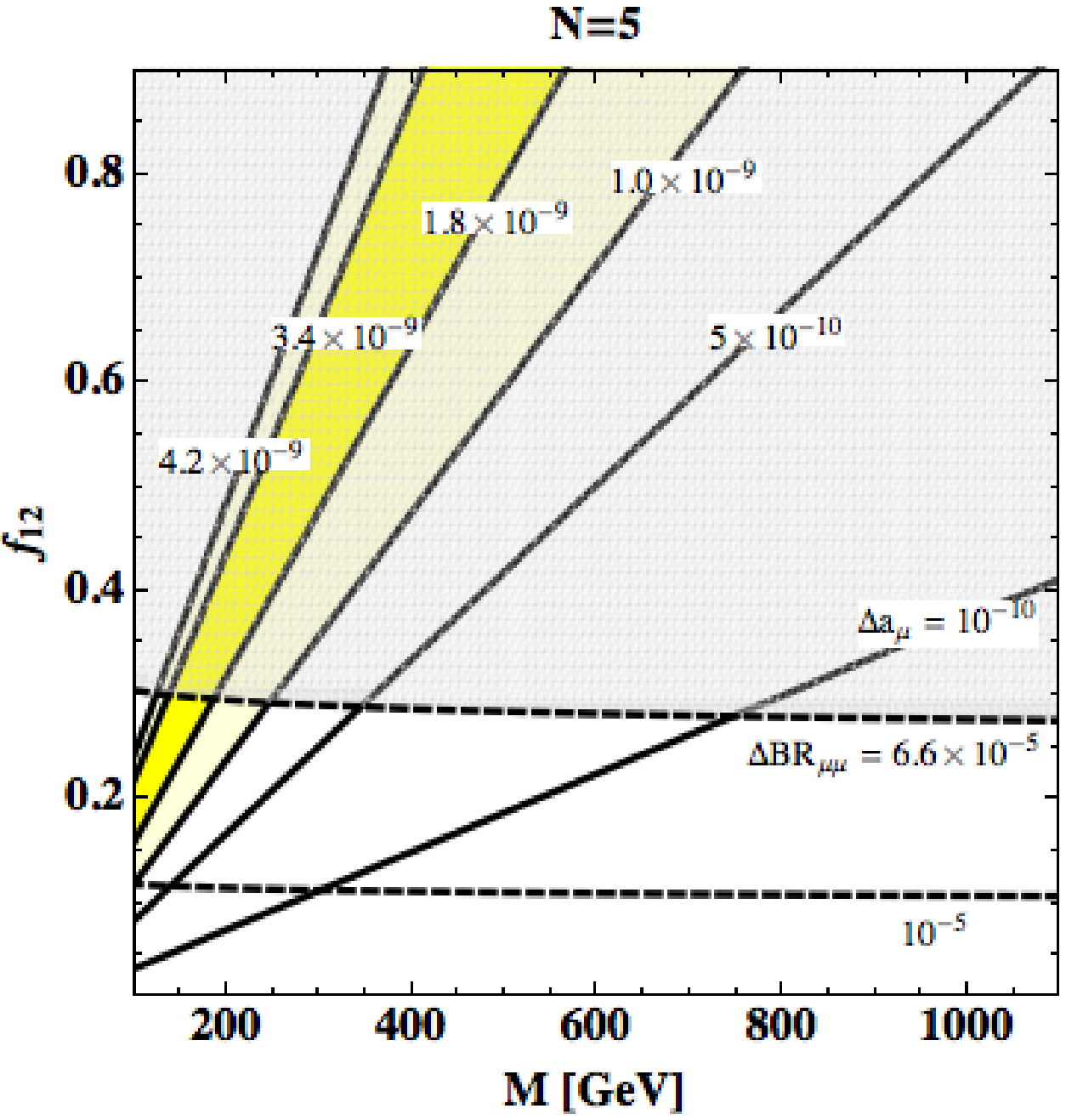}
 \includegraphics[width=7cm]{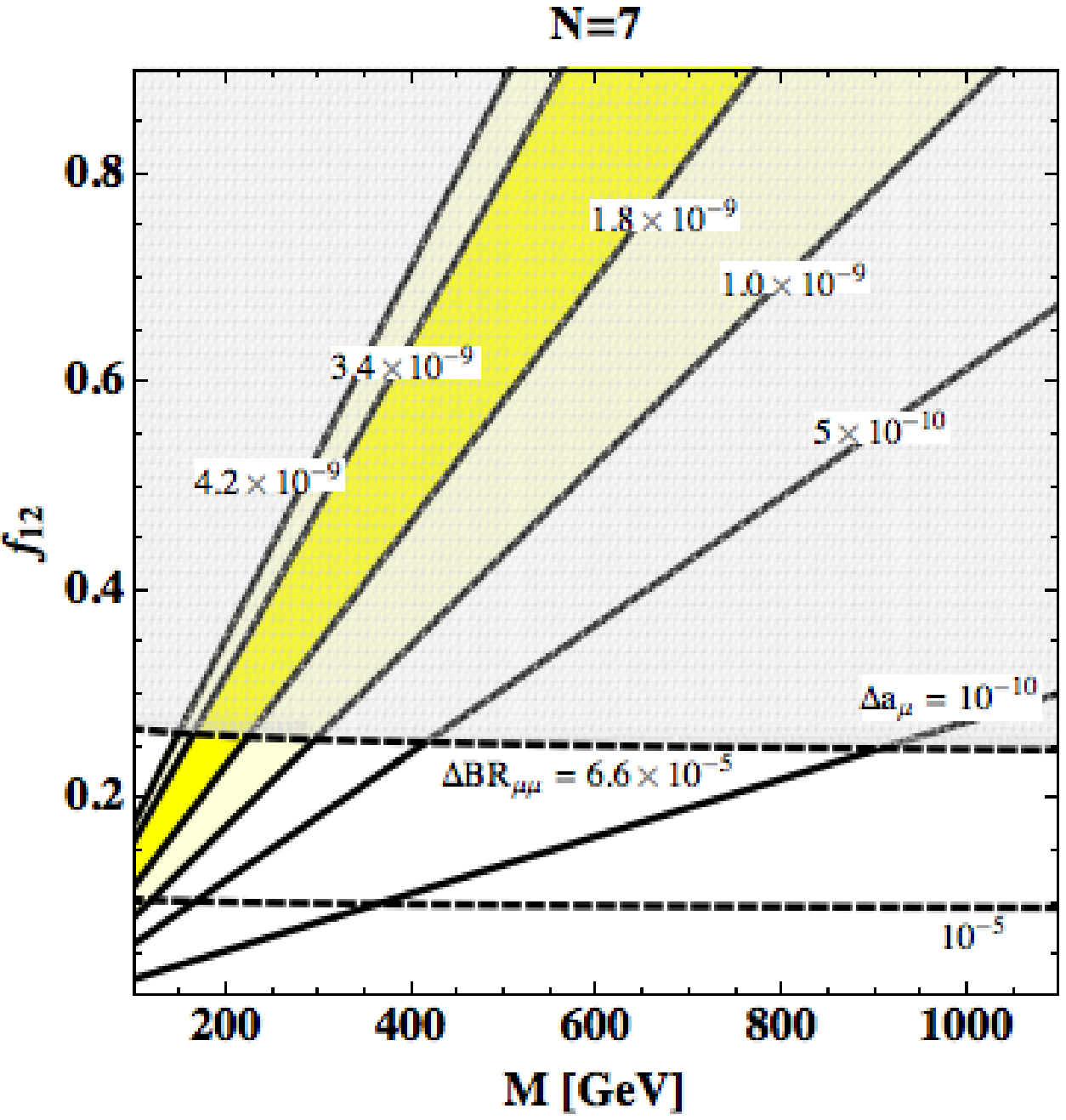}
\caption{Contours of $\Delta a_\mu$ and $\Delta BR_{\mu \mu} \equiv \Delta BR (Z \to \mu^+ \mu^-)$ on $\{M(=m_h = m_\psi), f_{21} \}$ plane where we take only $f_{21}$ to be non-zero and other $f_{ij}$ to be zero. 
The (light-)yellow region is $(2 \sigma) 1 \sigma$ region for muon $g-2$ and shaded region is excluded by $\Delta BR (Z \to \mu^+ \mu^-)$.}
\label{fig:contour}
\end{center}\end{figure}
  
  
\subsection{Collider physics and constraints}

In explaining muon $g-2$ by the interaction Eq.~(\ref{Eq:lag-yukawa}), the mass scale of exotic lepton doublet $L'$ is required to be 
less than $\sim 300$ GeV.
Thus exotic charged lepton can be produced at the LHC with sizable production cross section and we should take into account collider constraints 
to explore if the mass scale for explaining muon $g-2$ is allowed.
In our study, we focus on the exotic charged lepton with the highest electric charge since it has the largest pair production cross section and provide the most stringent constraint.

\begin{figure}[t!]
\begin{center}
\includegraphics[width=7cm]{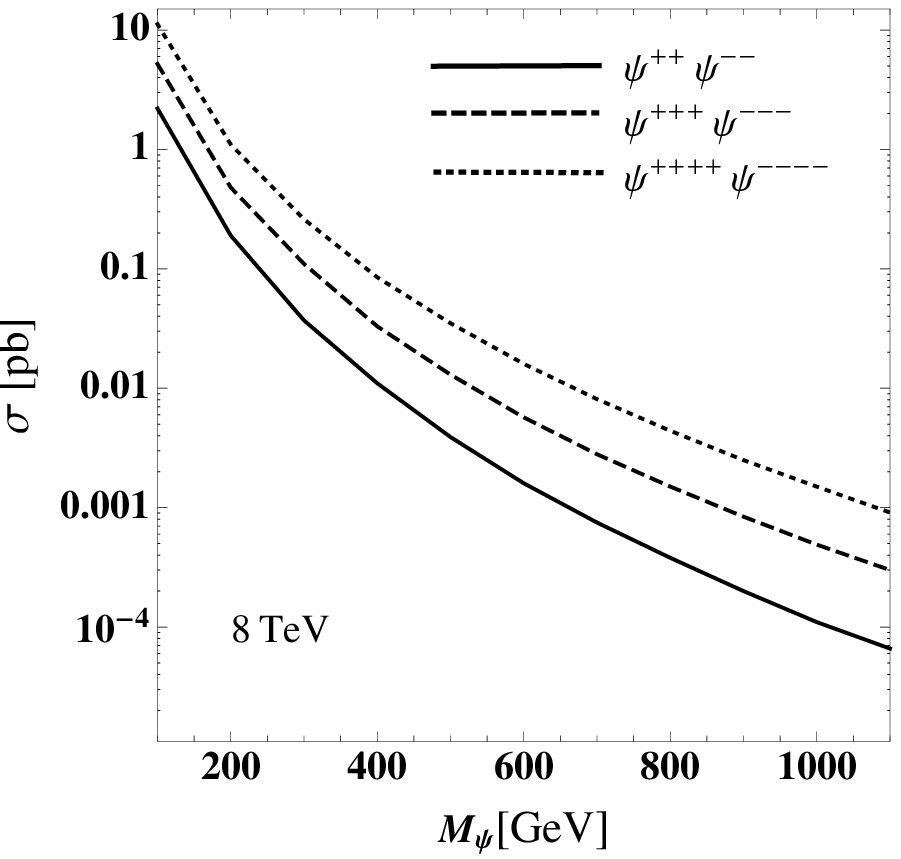}
 \includegraphics[width=7cm]{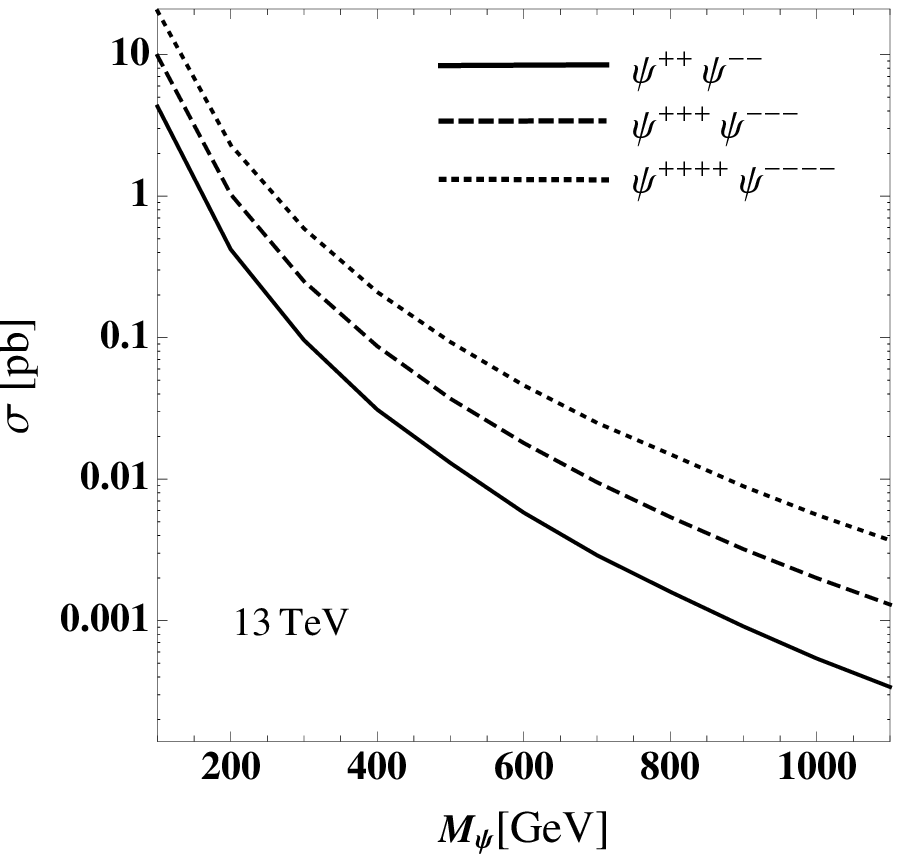}
\caption{The pair production cross section of the exotic charged leptons with the highest electric charged for each case at the LHC 8(13) TeV for left(right) plots.}
\label{fig:CX}
\end{center}\end{figure}

Firstly, we estimate the pair production cross section of the highest charged leptons for each case.
These charged leptons can be pair produced by Drell-Yan(DY) process, $q \bar q \to Z/\gamma \to \psi^{+n} \psi^{-n}$, and also by photon fusion(PF) process $\gamma \gamma \to \psi^{+n}  \psi^{-n}$~\cite{Babu:2016rcr, Ghosh:2017jbw, Ghosh:2018drw}. 
Here we estimate the cross section applying {\tt MADGRAPH/MADEVENT\,5}~\cite{Alwall:2014hca}, where the necessary Feynman rules and relevant parameters of the model are implemented using FeynRules 2.0 \cite{Alloul:2013bka} and the {\tt NNPDF23LO1} PDF~\cite{Deans:2013mha} is adopted.
In Fig.~\ref{fig:CX} we show the cross sections including both DY and PH processes at the LHC 8(13) TeV for left(right) plots.
We thus find that cross section is large when electric charge is increased where PF process highly enhance the cross section.

Secondly we list the decay chain of the highest charged lepton for each case. \\
{\bf (1) $N =3$}:  The decay chain of $E^{\pm \pm}$ is 
\begin{equation}
\psi^{\pm \pm} \to \ell^\pm_i h^{\pm (*)} \to \ell^\pm_i \ell^\pm_j \nu, 
\end{equation}
where charged scalar can be ether on-shell or off-shell.
Thus, $\psi^{++} \psi^{--}$ pair production process gives four charged leptons with missing transverse energy.
The singly charged scalar with $m_{h^+} > 100$ GeV is allowed by collider experiment and we require the mass is heavier than 100 GeV~\cite{pdg}.  \\
{\bf (2) $N =5$}:  The decay chain of $\psi^{\pm \pm \pm}$ is 
\begin{equation}
\psi^{\pm \pm \pm} \to \ell^\pm_i h^{\pm \pm (*)}  \to \ell^\pm_i \ell^\pm_j  \ell^\pm_k \left[\to \ell_i^\pm h^\pm h^\pm \to \ell^\pm_i \ell^\pm_j \ell^\pm_k \nu \right], 
\end{equation}
where charged scalar can be ether on-shell or off-shell as previous case, and process in square bracket can be induced introducing singly charged scalar with interaction Eq.~(\ref{EQ:extra_int_1}).
Thus, $\psi^{+++} \psi^{---}$ pair production process gives six charged leptons. 
Note that doubly charged scalar mass is constrained by the LHC data as $m_{h^{\pm \pm}} \gtrsim 700-800$ GeV and $m_{h^{\pm \pm}} \gtrsim 400$ GeV when $h^{\pm \pm}$ decay into $e^\pm e^\pm (\mu^\pm \mu^\pm)$ and $\tau^\pm \tau^\pm$ respectively~\cite{CMS:2017pet, Aaboud:2017qph}. 
The constraint is looser as $m_{h^{\pm \pm} } \gtrsim 200$ GeV when $h^{\pm \pm}$ dominantly decay via $h^{\pm \pm} \to h^\pm h^\pm \to \ell^\pm_i \ell^\pm_j \nu \nu$ process. 
To explain muon $g-2$, we require $h^{\pm \pm}$ to dominantly decay into singly charged scalars~\cite{Primulando:2019}. \\
{\bf (3) $N =7$}:  The decay chain of $\psi^{\pm \pm \pm \pm}$ is 
\begin{equation}
\psi^{\pm \pm \pm \pm} \to \ell^\pm_i h^{\pm \pm \pm (*)} \to \ell^\pm_i h^{\pm \pm (*)} h^{\pm (*)} \to h^{\pm (*)} h^{\pm (*)} h^{\pm (*)}  \to \ell^\pm_i \ell^\pm_j  \ell^\pm_k \ell^\pm_l \nu \nu \nu, 
\end{equation}
where triply charged scalar decays via interaction in Eq.~(\ref{EQ:extra_int_3}).
Also as in the previous case, we require doubly charged scalar decay into same sign singly charged scalar pair.
Thus, $\psi^{++++} \psi^{----}$ pair production process gives {eight} charged leptons with missing transverse energy.
In general, constraint on mass of triply charged scalar is weaker than that on $\psi^{\pm \pm \pm \pm}$ and we will not explicitly discuss the constraint.

Finally, let us discuss collider constraints on our scenario to explain muon $g-2$.
We note that the highest charged lepton dominantly decay into $\psi^{\pm n} \to \mu^\pm h^{\pm n-1}$ since $f_{21}$ coupling is required to be large for explaining muon $g-2$.
In addition to the conditions discussed above we classify benchmark scenarios as follows: \\
{\bf (a)} singly charged scalar decay into $\ell = e, \mu$ in decay chain and exotic charged lepton has sufficiently short decay length, \\
{\bf (b)} exotic charged leptons have long decay length and pass through detector, \\
{\bf (c)} singly charged scalar decays into $\tau \nu$ mode and the highest charged scalar mass is slightly lighter than that of the highest charged lepton. 

\begin{figure}[t!]
\begin{center}
\includegraphics[width=7cm]{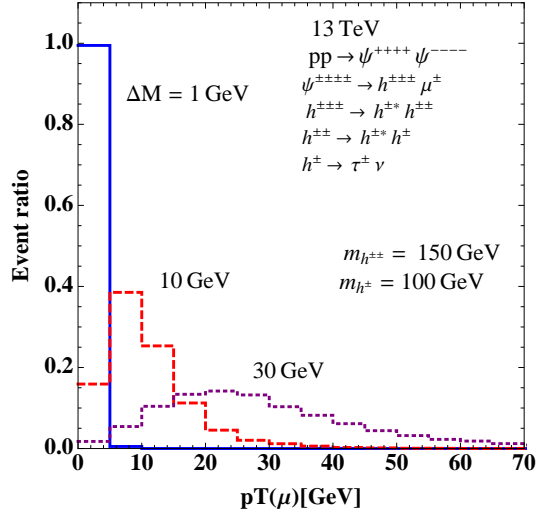}
\caption{The distribution of transverse momentum of muon $\mu$ in $\psi^{\pm \pm \pm \pm} \to h^{\pm \pm \pm} \mu^\pm \to h^{\pm \pm} \mu^\pm \tau^\pm \nu \to h^\pm \mu^\pm \tau^\pm \tau^\pm \nu \nu \to \mu^\pm \tau^\pm \tau^\pm \tau^\pm \nu \nu \nu$ decay at the LHC 13 TeV (neutrino and anti-neutrino are not distinguished here); the vertical axis shows event ration defined by $N_{\rm bin}/N_{\rm total}$ where $N_{\rm total}$ and $N_{\rm bin}$ indicate number of events in total and those inside corresponding bins. Here {$\Delta M$} indicate mass difference between $\psi^{\pm \pm \pm \pm}$ and $h^{\pm \pm \pm}$.}
\label{fig:distPT}
\end{center}\end{figure}

\begin{figure}[t!]
\begin{center}
\includegraphics[width=5cm]{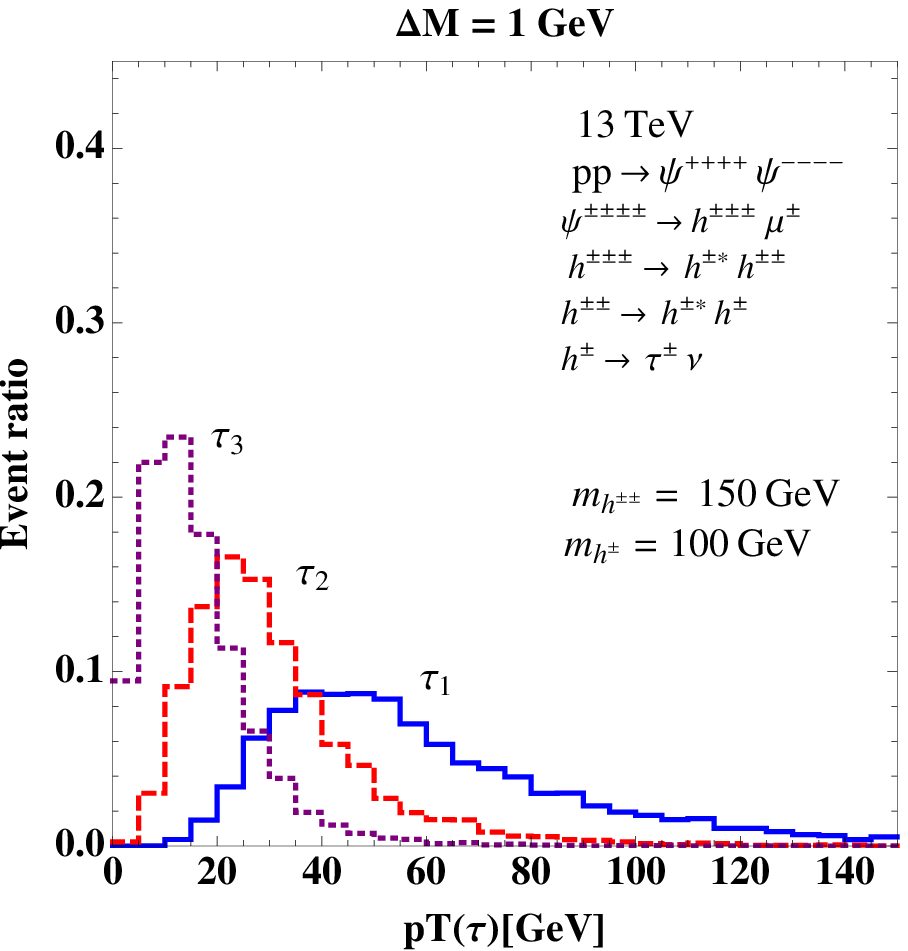}
\includegraphics[width=5cm]{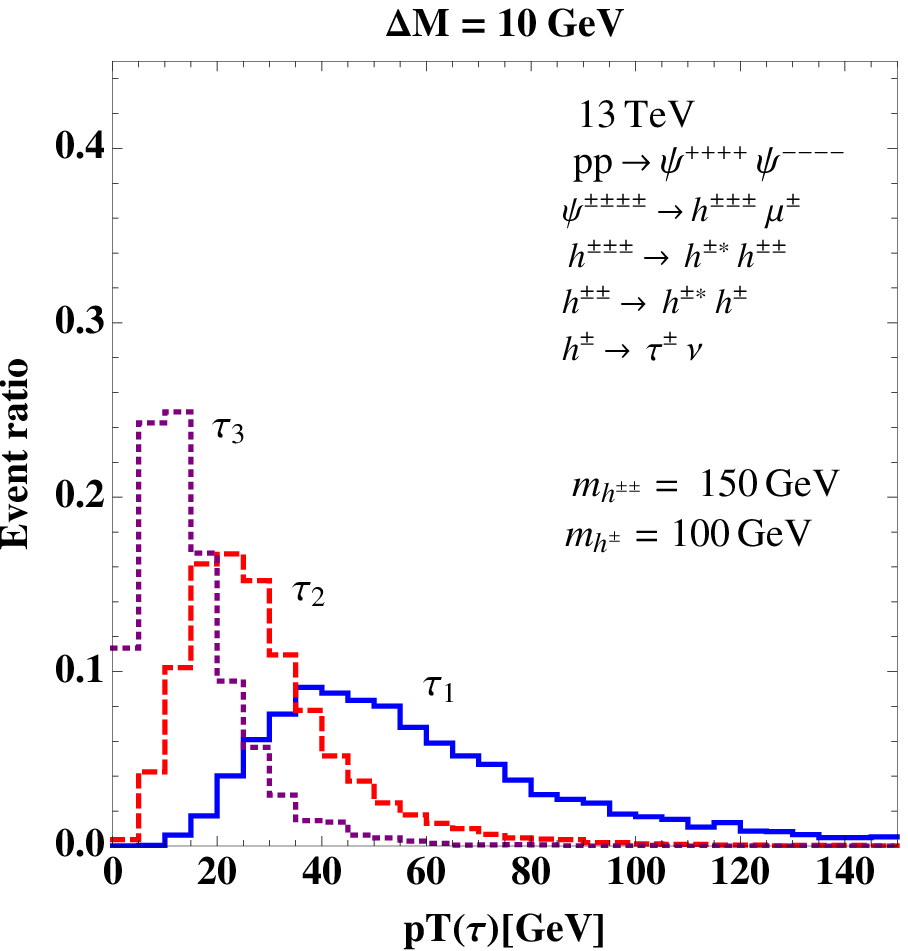}
\includegraphics[width=5cm]{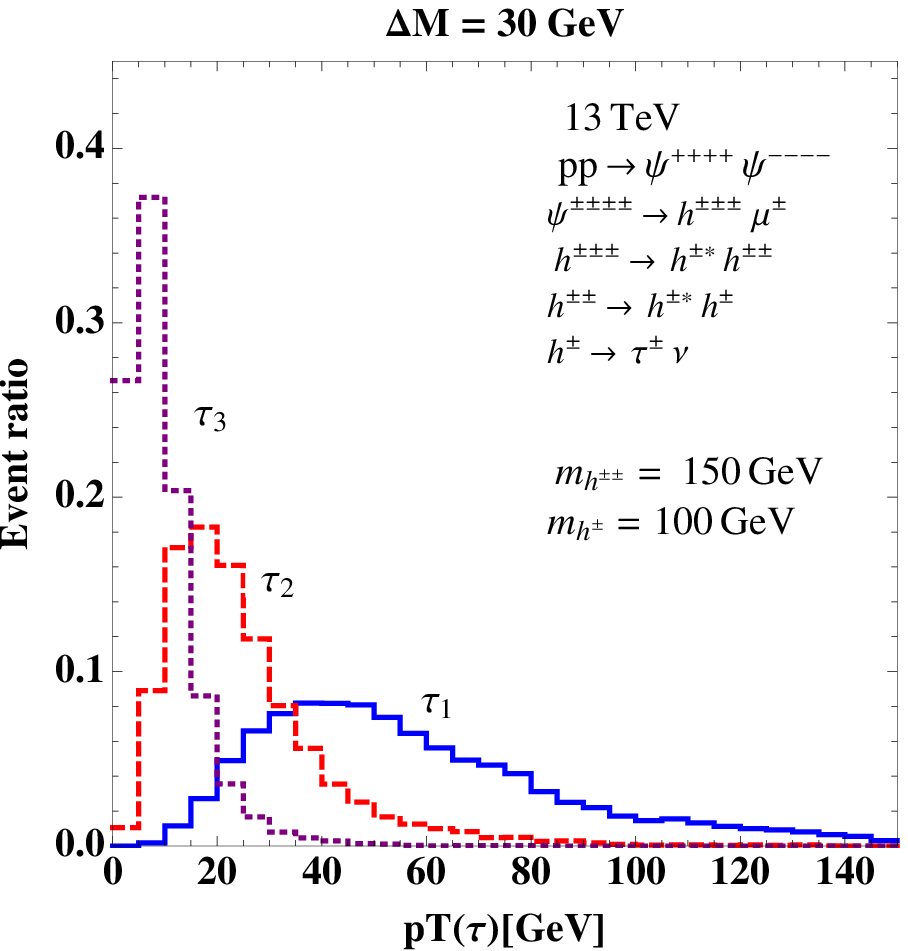}
\caption{The distribution of transverse momentum of muon $\tau$ leptons for the same process as Fig.~\ref{fig:distPT} where three $\tau$ leptons are distinguished by transverse momentum as $p_T(\tau_3) < p_T(\tau_2) < p_T(\tau_1)$ for each events. }
\label{fig:distPT2}
\end{center}\end{figure}

For scenario {\bf (a)}, inclusive multi-lepton search constrains the cross section where upper bound of the cross section is $\sim 1$ fb at the LHC 8 TeV for the signal in which number of charged lepton $N_\ell$ ($\ell = e, \mu$) is $N_{\ell} \geqslant 3$~\cite{Aad:2014hja}.
Comparing the cross section for 8 TeV in Fig.~\ref{fig:CX}, the charged lepton masses are required to be $m_\psi \gtrsim (650, 900, 1100)$ GeV.
In this scenario, the region explaining muon $g-2$ in 1$\sigma$ is excluded for all $N$ and the largest value of muon $g-2$ is roughly $\Delta a_\mu \sim 10^{-10}$ for each case.
Scenario {\bf (b)} can be realized when charged scalar in decay chains is off-shell and extra couplings in Eq.~(\ref{EQ:extra_int_1})-(\ref{EQ:extra_int_3}) are sufficiently small.
For long-lived charged particle, upper bound of the cross section is given in ref.~\cite{Aaboud:2019trc} for the LHC 13 TeV.
Comparing the result for chargino, we find the upper limit is less than $1$fb, and since we have multiply charged leptons the constraint will be stronger. 
Thus the collider constraint in this scenario is stronger than the scenario (a) and we cannot expect sizable muon $g-2$.
For scenario {\bf (c)}, the decay chain provides signature for each case such that
case (1) gives low energy muon with missing transverse energy, and case (2) and (3) give multi-tau lepton signature with low energy muon since we require mass difference between $\psi^{-n-1}$ and $h^{\pm n}$ is small and $h^{\pm n}$ is on-shell.
{In Fig.~\ref{fig:distPT}, we show the event ratio for the distribution of transverse momentum of muon, $\mu$, in $\psi^{\pm \pm \pm \pm} \to h^{\pm \pm \pm} \mu^\pm \to h^{\pm \pm} \mu^\pm \tau^\pm \nu \to h^\pm \mu^\pm \tau^\pm \tau^\pm \nu \nu \to \mu^\pm \tau^\pm \tau^\pm \tau^\pm \nu \nu \nu$ decay chain at the LHC 13 TeV 
for different values of {$\Delta M$} indicating mass difference between $\psi^{\pm \pm \pm \pm}$ and $h^{\pm \pm \pm}$
where the behaviors are similar if we change colliding energy from 13 TeV to 8 TeV or 14 TeV; here we consider case (3) but we will have similar results for the other cases.
The masses of $h^{\pm \pm}$ and $h^{\pm}$ are also fixed to be $m_{h^{\pm \pm}} = 150$ GeV and $m_{h^\pm} = 100$ GeV.
In Fig.~\ref{fig:distPT2}, we also show the event ratios for the distribution of transverse momentum of $\tau$ in the same process where three $\tau$ leptons are distinguished by transverse momentum as $p_T(\tau_3) < p_T(\tau_2) < p_T(\tau_1)$ for each event.
It is found that transverse momentum of some $\tau$ leptons are generally sizable and they can be detected at detector.
On the other hand transverse momentum of $\mu$ tends to be small for $\Delta M \lesssim 10$ GeV and it will be missed by event trigger.
For multi-lepton search in ref.~\cite{Aad:2014hja}, they require one muon or electron should have $p_T > 26$ GeV and $p_T > 15$ GeV from the second muon(electron).
As a result number of dimuon signal events becomes less than $\sim 0.1 \%$ after trigger for $\Delta M = 10$ GeV.
We thus see that if $\Delta M \lesssim 10$ GeV most of events are missed by event trigger and we can escape experimental bound.
Therefore the scenario {\bf (c)} with small $\Delta M$ still can be allowed since analysis of multi-tau signature is more difficult and explicit bound is not given.}
Thus we conclude that to obtain sizable muon $g-2$ by interaction Eq.~(\ref{Eq:lag-yukawa}) we should rely on this specific scenario.
Therefore multi-tau signature is important to test the mechanism to explain muon $g-2$ although analysis of it is challenging.

\section{Conclusions}

We have analyzed muon $g-2$, LFVs, and $Z$ decays including collider physics in multi-charged particles.
 We have found LFVs do not restrict the allowed region of muon $g-2$, while $Z\to\nu_i\bar\nu_j$ invisible decay and $Z\to\mu\bar\mu$
 give stringent constraints and the allowed region is drastically disappeared. Also, larger $N$ increases the allowed region of muon $g-2$. 
{\it However once we consider the constraint of collider physics, the typical size of muon $g-2$ is of the order $10^{-10}$},
depending on the benchmark scenarios in (a,b,c).
To obtain sizable muon $g-2$ of $\mathcal{O}(10^{-9})$, we have found that the specific scenario is required for decay chain of the charged particles 
in which the mass of $L'$ is slightly heavier than $h^{\pm n}$ and charged scalar bosons decay into mode only including $\tau$ and neutrinos.
Therefore analysis of multi-tau lepton signature is important to fully test the scenario to explain muon $g-2$.

\section*{Acknowledgments}
This research was supported by an appointment to the JRG Program at the APCTP through the Science and Technology Promotion Fund and Lottery Fund of the Korean Government. This was also supported by the Korean Local Governments - Gyeongsangbuk-do Province and Pohang City (H.O.). H. O. is sincerely grateful for KIAS and all the members.


\end{document}